\documentclass[%
 aip,
 amsmath,amssymb,
 reprint,%
]{revtex4-1}

\usepackage{graphicx}
\usepackage{dcolumn}
\usepackage{bm}

\usepackage[T1]{fontenc}
\usepackage{mathptmx}

\begin{document}

\preprint{AIP/123-QED}

\title{Statistical screening model for moderately coupled and dense plasmas}

\author{Fuyang Zhou}
    \affiliation{Key Laboratory of Computational Physics, Institute of Applied Physics and Computational Mathematics, Beijing,100088, China}
\author{Yizhi Qu}
    \affiliation{School of Optoelectronics, University of Chinese Academy of Sciences, Beijing, 100049, China}
\author{Junwen Gao}
    \affiliation{Laser Fusion Research Center, China Academy Of Engineering Physics, Mianyang,621900, China }
\author{Yulong Ma}
    \affiliation{Key Laboratory of Computational Physics, Institute of Applied Physics and Computational Mathematics, Beijing,100088, China}
\author{Yong Wu}
    \email[Correspondence email address: ]{wu_yong@iapcm.ac.cn}
    \affiliation{Key Laboratory of Computational Physics, Institute of Applied Physics and Computational Mathematics, Beijing,100088, China}
\author{Jianguo Wang}
    \affiliation{Key Laboratory of Computational Physics, Institute of Applied Physics and Computational Mathematics, Beijing,100088, China}

\date{\today} 

\begin{abstract}
For atoms embedded in dense plasma, the plasma screening effects will greatly alter their structure and dynamics, and then determine the radiation transport properties of the plasma. 
In the present work, a new statistical model is proposed for treating electron screening effects on atoms in moderately/strongly coupled and dense plasmas, in which the three-body processes are found to significantly influence the plasma-electron density distributions and leads to a dependence of the distribution on the specific bound state of the targeted atom. 
As a critical check, the model is applied to simulate the emission spectra of He-like Aluminum and Chlorine in hot dense plasmas, and much better agreements of the line shifts are obtained with the experiments of Stillman \textit{et al.} in 2017 and Beiersdorfer \textit{et al.} in 2019 than previous calculation results. 
Compared with the classical molecular dynamic simulations of electron distributions in moderately coupled plasmas, the present model can better describe the low-energy electron distribution than other models and the multi-body effects are well considered. 
The present model provides a promising tool to reasonably treat the electron screening effect of non-equilibrium dense plasma on atoms with specific bound states, which is urgently needed in high-precision simulations of the atomic processes, plasma spectra and radiation transport properties etc. \end{abstract}


\maketitle

Warm/hot dense plasma exists widely in all types of stars \cite{Tayler}, the interior of giant planets \cite{Helled} and inertial confinement fusions \cite{Lindl,S.X.Hu}. 
How to understand the microscopic and macroscopic properties of such plasmas in extreme conditions are of great importance, and extensive experiments have been performed to study its thermodynamic and transport properties in the last decade \cite{Vinko,Ciricosta1,Ciricosta2,Hurricane,Gomez,Bailey,Nagayama}. 
For atoms embedded in dense plasma, electron screening significantly affects the atomic energy levels and wave functions, resulting in ionization potential depression (IPD) and line shift, and greatly influence all photon, electron and ion scattering processes and corresponding cross sections \cite{janev,ma2020}.
As the integrated parameters of fundamental atomic processes data, the resultant electron screening effects on spectra (opacity) and equation of state (EOS) of dense plasmas would be notable and should be carefully taken into consideration \cite{Salzmann,Griem}. 
Reversely, the screening effect on plasma spectra can be used as a powerful tool for diagnosing electron density and temperature of plasmas \cite{Griem,gomez2020,duan2014}. 

In order to describe the plasma screening effects, various models have been proposed since the pioneering work of Debye and H\" uckel \cite{Debye} in 1920s, which is valid for the case of non-degenerate and weakly coupled plasmas. 
Thereafter, the Thomas-Fermi (TF) screening and self-consistent field ion-sphere (IS) models have been developed to study the electron screening effects in degenerate plasmas \cite{Ichimaru,Zerah}. 
For the convenience of application, various analytic models, such as the ones of uniform electron gas (UEGM) \cite{Crowley}, Ecker-Kröll (EK) \cite{EK}, Stewart-Pyatt (SP) \cite{SP}, and analytic fits to ion-sphere potentials \cite{Rosmej,Lixiangdong,Lixiangdong2}, are then proposed and widely used to calculate the IPDs and line shifts in dense plasmas. 
However, these models are constructed with different approximations of electron distribution based on Boltzmann or Fermi-Dirac distributions, and are found unreliable in latest high-precision spectra experiments for warm/hot dense plasmas, such as the measurements of IPDs \cite{Ciricosta1,Ciricosta2,Hoarty,Fletcher} and line shifts \cite{Stillman,Beiersdorfer}.

With the great advances in obtaining uniform, well-characterized and high-energy-density plasmas, it becomes possible to precisely measure the ionic level shift of warm/hot dense plasmas, which can be used to benchmark the screening models \cite{Vinko,Hoarty,Fletcher,Stillman}. 
Stillman \textit{et al.} \cite{Stillman} and Beiersdorfer \textit{et al.} \cite{Beiersdorfer} recently performed line-shift measurements of Al$^{11+}$ and Cl$^{15+}$ in hot-dense plasmas, and obvious disagreements are found in comparison with the predictions of the Li and Rosmej’s analytical model \cite{Lixiangdong}, which is a fit of the self-consistent field ion-sphere model. 
The significant discrepancies between measurements and calculations attract lots of attentions, and various models are applied to handle this issue, including numerical self-consistent field ion-sphere models \cite{Lixiangdong3,chenzhanbin1,chenzhanbin2}, line-shift model based on Stewart-Pyatt screening potential \cite{Gu} and other analytical fits of ion-sphere models \cite{Lixiangdong3,Iglesias,Singh}.
These screening models are found only work for $1s3p-1s^2$ transition of Cl$^{15+}$ but failed to the case of $1s2p-1s^2$ transition of Al$^{11+}$ in dense plasmas, which indicates that a reliable model is needed to distinguish the different screening effects on different transitions. 
In this letter, a new statistical model is proposed for treating electron screening effects in moderately/strongly coupled and dense plasmas, in which the effects of three-body processes on the electron distribution are considered and leads to different electron screening effects on different transitions. 
As a critical check, the present model is applied to compute the line shifts of Al$^{11+} (1s2p-1s^2)$ and  Cl$^{15+} (1s3p-1s^2)$ transitions of the latest two experiments \cite{Stillman,Beiersdorfer}.

\textbf{\textit{Electron screening effect on targeted ions}} $-$ 
The plasma-electron density distribution is the kernel of an electron screening model. For the well-known Fermi-Dirac distribution
\begin{equation}\label{eq1}
\ f_{FD}(p,\boldsymbol{r})=\frac{1}{1+exp\left[\frac{1}{k_B T}\left(\frac{p^2}{2m_e}-e\Phi(\boldsymbol{r})-\mu \right)\right]},
\end{equation} 
the number density of electrons is given as
\begin{equation}\label{eq2}
\rho(\boldsymbol{r})=\frac{1}{2\pi^2\hbar^3}\int_{0}^{\infty} f_{FD}(p,\boldsymbol{r})p^2 dp
\end{equation} 
where $\mu$ is the chemical potential, $\Phi(\boldsymbol{r})$ is the total effective potential at position $\boldsymbol{r}$, and $p$ is the magnitude of electron momentum. 
In UEGM and self-consistent field ion-sphere models the Fermi-Dirac distribution is applied, but only consider the screening effects of the free electrons, which are defined by the condition
\begin{equation}\label{eq3}
p>p_0=\sqrt{2m_e e\Phi(\boldsymbol{r})},
\end{equation} 
guaranteeing their kinetic energies are larger than the absolute values of their potential energies. 

\begin{figure}[htbp]
\centering
\includegraphics[width=5cm]{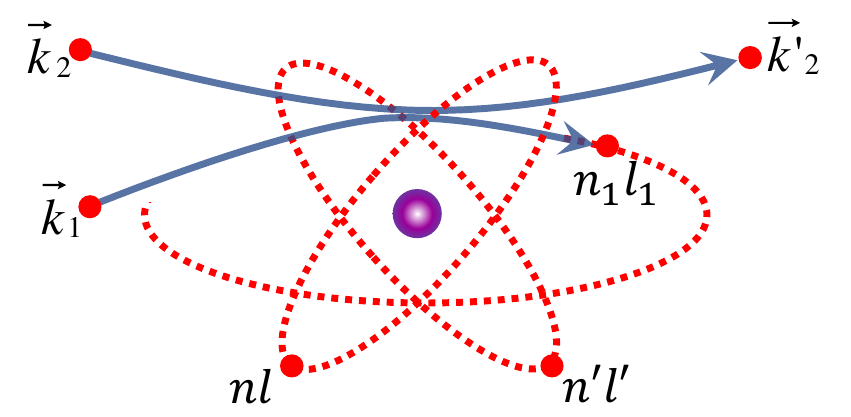}
\caption{The scheme of the three-body recombination process between free electrons and the targeted ion with specific bound state. Noted that the initial bound-state orbitals are fixed, and the recombined bound electrons produce the negative-energy electron distribution around the targeted ion.}\label{fig0}
\end{figure} 

However, for the dense plasmas the three-body recombination processes and its effect on electron distribution become important, as shown in figure \ref{fig0}. 
Throughout this work, the recombined bound electrons are treated as negative-energy ones. 
For an ideal plasma, in which the mean kinetic energy of an electron is much larger than its mean potential energy, free-electron distribution dominates the total electron density distribution and the contribution of negative-energy electrons is negligible. 
But for non-ideal plasmas, such as moderately/strongly coupled plasmas, the distribution of negative-energy electron is expected to significantly influence the total electron density distribution around the specific ion, and will deviate from the electron density distribution in equilibrium. 
For example, the IPDs and line shifts arise from the electron screening effect on the specific bound state of the ions, which cannot be handled by the Fermi-Dirac distribution of the equilibrated plasma system consisting of free electrons, ions and atoms. 
When investigating electron screening effect on an ion in $nln'l'$ state, the contribution of itself and the transitions involved $nln'l'$ state on electron density distribution should not be included. 
In this case, the negative-energy electron distribution will deviate from the equilibrated distribution and should be retrieved.

\textbf{\textit{Model formulation}} $-$ 
In steady-state approximation, the distribution of negative-energy state keeps stable at given free-electron temperature and density. 
For simplicity, we consider a plasma consisting of free electrons and singly charged ions with number density $n_e$, $n_{ion}$. The three-body recombination process leads to distribution of negative-energy electrons, which are described as atoms in electronic state $j$ with the population density $n_{atom}^j$.The rate equation of $n_{atom}^j$ meets with 
\begin{equation}\label{eq4}
\begin{array}{c}
\frac{dn_{atom}^j}{dt}=\sum_{j^\prime}{{(n}_en_{atom}^{j^\prime}K_{j^\prime j}-n_en_{atom}^jK_{{jj}^\prime})}\\
+n_en_{atom}^j\alpha_j-n_e^2n_{ion}\beta_j
\end{array}
\end{equation} 
\begin{equation}\label{eq5}
\text{and}\ \  n_{atom}^j=\frac{n_e^2n_{ion}\beta_j+\sum_{j^\prime}{n_en_{atom}^{j^\prime}K_{j^\prime j}}}{n_e\alpha_j+\sum_{j^\prime}{n_eK_{{jj}^\prime}}}.
\end{equation}
Here, $\alpha_j$ and $\beta_j$ are the rate coefficients of electron impact ionization and three-body recombination, $K_{j^\prime j}$ and $K_{{jj}^\prime}$ are the rate ones of excitation and de-excitation.
For an equilibrated plasma with same $n_e$ and $n_i$, the detailed balance relations are met as
\begin{equation}\label{eq6}
\begin{array}{c}
n_en_{atom,equil}^{j^\prime}K_{j^\prime j}=n_en_{atom,equil}^jK_{{jj}^\prime} 
\\
\text{and}\ \ {n_en}_{atom,equil}^j\alpha_j=n_e^2n_{ion}\beta_j
\end{array},
\end{equation}
where $n_{atom,equil}^j$ can be well described using Fermi-Dirac statistics theory.

Combined with equations (\ref{eq5}) and (\ref{eq6}), the $n_{atom}^j$ can be written as
\begin{equation}\label{eq7}
\begin{array}{l}
n_{atom}^j=n_{atom,equil}^j\frac{n_e^2n_{ion}\beta_j+\sum_{j^\prime}{n_en_{atom}^{j^\prime}K_{j^\prime j}}}{n_e^2n_{ion}\beta_j+\sum_{j^\prime}{n_en_{atom,equil}^{j^\prime}K_{j^\prime j}}}\\
=n_{atom,equil}^j \chi^j
\end{array}.
\end{equation}
Due to that the contributions of the studied bound state and the transitions involved it are not included, there is $\sum_{j^\prime} n_{atom}^{j^\prime} < \sum_{j^\prime} n_{atom,equil}^{j^\prime}$, and $\chi^j$ can be approximated as
\begin{equation}\label{eq8}
\chi^j=\frac{n_e^2n_{ion}\beta_j}{n_e^2n_{ion}\beta_j+\sum_{j^\prime}{n_en_{atom,equil}^{j^\prime}K_{j^\prime j}}},
\end{equation}
which is defined as the non-equilibrium coefficient. Combining the classical scattering theory and equilibrium distribution theory, the non-equilibrium coefficient can be derived as
\begin{equation}\label{eq9}
\begin{array}{c}
\chi\left(\boldsymbol{p},\boldsymbol{r}\right)=2\sqrt{\frac{\varepsilon_r}{{\pi k}_BT}}exp\left(\frac{\varepsilon_r}{{ k}_BT}\right)/Erf\left(\sqrt{\frac{\varepsilon_r}{{ k}_BT}}\right)
\end{array},
\end{equation} 
where $Erf\left(x\right)$ is the error function and $\varepsilon_r=e\Phi\left(\boldsymbol{r}\right)-p^2/(2m_e)$. 
The detail derivation of Eq. (\ref{eq9}) is presented in supplemental material \cite{SI}. Then, the total plasma-electron density $\rho\left(\boldsymbol{r}\right)$ is given by
\begin{equation}\label{eq10}
\begin{array}{c}
\rho\left(\boldsymbol{r}\right)=\frac{1}{2\pi^2\hbar^3}[\int_{\sqrt{2m_e(\varepsilon_b-e\Phi(\boldsymbol{r}))}}^{p_0}{f_{FD}\left(p,\boldsymbol{r}\right)\chi\left(p,\boldsymbol{r}\right)p^2dp}
\\
+\int_{p_0}^{\infty}{f_{FD}\left(p,\boldsymbol{r}\right)p^2dp}]
\end{array},
\end{equation} 
in which the effective potential $\Phi(\boldsymbol{r})$ is dependent on $\rho(\boldsymbol{r})$ and calculated though a self-consistent iteration process.
It is noted that this model can naturally converge to Debye-H\" uckel model at the weak-coupling limit and to ion-sphere model at the limit of strongly degeneracy, and it is same with the Fermi-Dirac distribution for the free-electron distribution. 

In equation (\ref{eq10}), the momentum of negative-energy electron is restricted be larger than $\sqrt{2m_e(\varepsilon_b-e\Phi(\boldsymbol{r}))}$, where $\varepsilon_b$ is the energy of outermost bound electron. 
This limitation is due to the degeneracy effect (the Pauli exclusive principle) between recombined and initial bound electrons. 
Meanwhile, the distribution of initial bound electrons will influence the total effective potential $\Phi(\boldsymbol{r})$ of a targeted ion
\begin{equation}\label{eq11}
\Phi(\boldsymbol{r})=\int \frac{e}{4\pi \varepsilon_0|\boldsymbol{r}-\boldsymbol{r}'|}\left[Z\delta(\boldsymbol{r}')-\rho_b(\boldsymbol{r}')-\delta\rho(\boldsymbol{r}')\right]d\boldsymbol{r}',
\end{equation}
\begin{equation}\label{eq12}
\text{and  }\delta \rho(\boldsymbol{r})=\rho(\boldsymbol{r})-\rho_e
\end{equation}
where $\delta \rho(\boldsymbol{r})$ is the plasma-electron density fluctuation induced by the ion, $\rho_e$ is the mean electron density, and the density of bound electrons $\rho_b (\boldsymbol{r})$  is calculated by
\begin{equation}\label{eq13}
\rho_b(\boldsymbol{r})=\frac{1}{4\pi}\sum_{j}q_j[P_j^2(r)+Q_j^2(r)].
\end{equation}
Here $q_j$ is the occupation number of electrons in the orbital $j$, and $P_j (r)$ and $Q_j (r)$ are the relativistic radial wave functions of the large and small components, respectively.
The bound wave functions are obtained by using multi-configuration Dirac-Fock approach \cite{MCDF1,MCDF2}.

Based on this model, self-consistent calculations of plasma screening effects on the atomic properties can be carried out without any adjustable parameter. The level shift for orbital $j$ can be computed by
\begin{equation}\label{eq14}
\begin{array}{c}
\Delta \varepsilon_j=\int_{0}^{\infty}[P_j^2(r)+Q_j^2(r)]V(r)r^2dr\\
\\
\text{with } V(r)=\int \frac{e^2}{4\pi\varepsilon_0|\boldsymbol{r}-\boldsymbol{r}'|}\delta\rho(\boldsymbol{r}')d\boldsymbol{r}'
\end{array},
\end{equation}
where $V(r)$ is the potential energy derived from plasma-electron distribution. 
Compared to the previous models, the contributions of recombined bound electrons to electron density distribution are considered in the present model, which is expected to be more applicable for non-equilibrium dense plasmas. 
In the following, the model is applied to calculate the line shifts observed in latest experiments \cite{Stillman,Beiersdorfer}.

\begin{figure}[htbp]
\centering
\includegraphics[height=4.1cm,width=8cm]{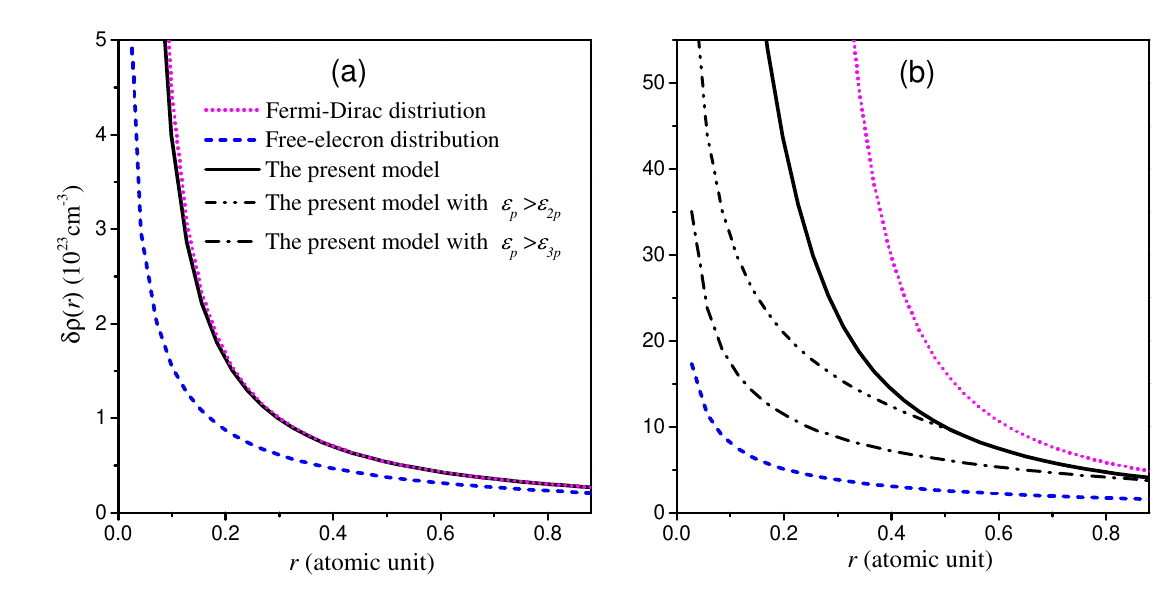}
\caption{ Comparison of electron density deviation $\delta \rho(r)$ induced by an Al ion (at origin) for different models, for ion charge of (a) $Q$=1, (b) $Q$=11.  The mean electron density $\rho_e$  is $3 \times 10^{23} \, /cm^3$, the temperature $T_e$  is 300 eV, and the orbital energies of $2p$ and $3p$ are about -498 eV and -217 eV, respectively. The distance $r$ is range from 0 to $a_{WS}/2\approx0.88$.}\label{fig1} 
\end{figure}

\textbf{\textit{Line-shift calculations}} $-$ 
In the line-shift experiments, the laser produced non-equilibrium plasmas are weakly coupled, with $(\rho_e, T_e)$=$(1-5\times10^{23} \, /cm^3 ,\  250-375 eV)$ for Al$^{11+}$ \cite{Stillman} and $(3-6\times10^{23} \, /cm^3, \  600-650 eV)$ for Cl$^{15+}$ \cite{Beiersdorfer}, respectively, for which the Coulomb coupling parameters $\Gamma_e=e^2/(4\pi\varepsilon _0 a_{WS} k_B T_e)$ are about 0.05 with $a_{WS}$ as the Wigner-Seitz radius.
Then, the present model is employed to obtain the electron distributions $\delta \rho(\boldsymbol{r})$ of Al$^{11+}$ and Cl$^{15+}$ with the above plasma conditions. 
In order to reveal the dependence of the charge state number $Q$, the $\delta \rho(\boldsymbol{r})$ for single charged ion ($Q$=1) is also computed and compared with the ones of Al$^{11+}$ ($Q$=11) and Cl$^{15+}$ ($Q$=15). 
It is found that the present results are consistent with those of Fermi-Dirac distribution for $Q$=1, but there is large difference between the present model and Fermi-Dirac distribution for Al$^{11+}$ and Cl$^{15+}$ in the weakly coupled plasma with $\Gamma_e \approx 0.05$. 
As a example, the comparisons between the present model and Fermi-Dirac distribution for $Q$=1 and $Q$=11 within the same conditions of $\rho_e=3\times10^{23}\, /cm^3,\,T_e=300$ eV and $\Gamma_e=0.052$ are presented in figure \ref{fig1} (a) and (b), respectively.
For the condition of $e\Phi\left(r\right) \approx Qe^2/(4\pi\epsilon_0\boldsymbol{r}) <k_BT$ the distribution of the present model is reduced to Fermi-Dirac one, for $e\Phi\left(r\right) >k_BT$ there is large difference at the low energy electron distribution, namely the negative-energy electron distribution, which is due to the contribution of three-body recombination processes, for details please see the discussion of figure 3 in the supplemental material \cite{SI}.

\begin{figure}[htbp]
\centering
\includegraphics[height=5.66cm,width=8cm]{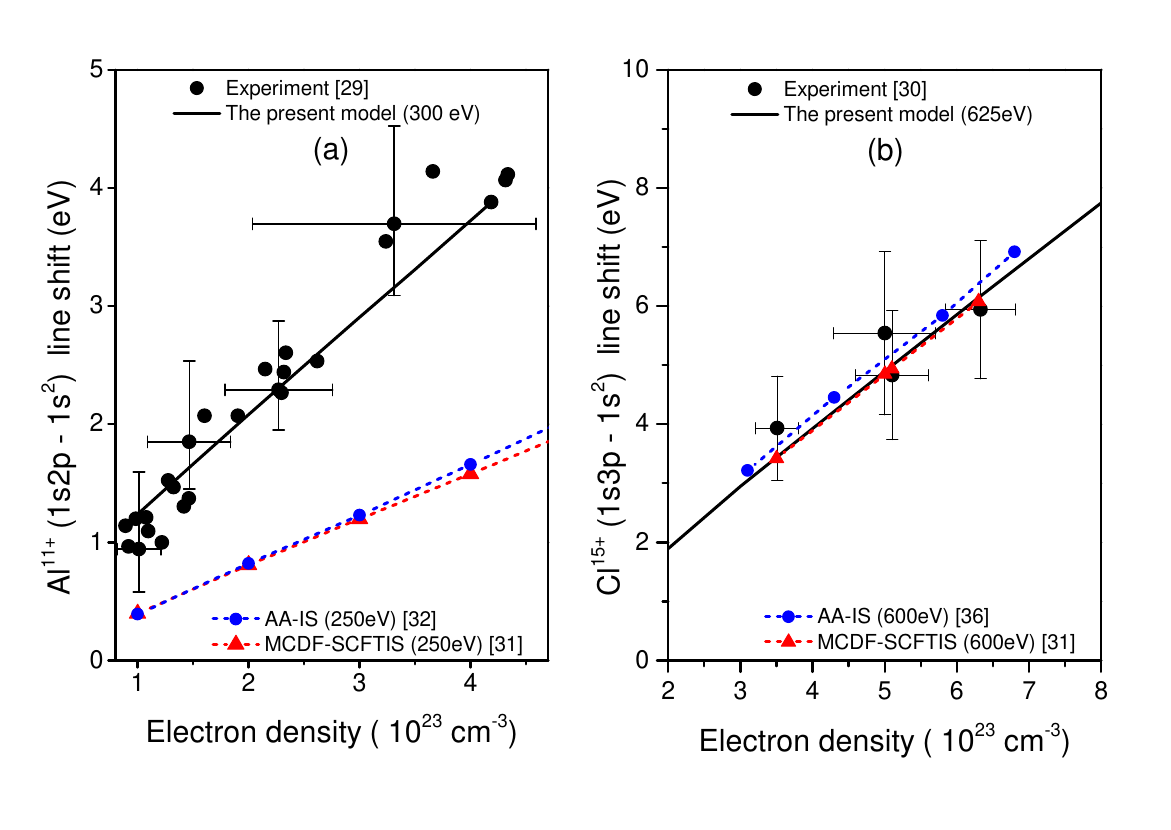}
\caption{ Comparison of The the line shifts from experimental measurements \cite{Stillman,Beiersdorfer} and the present screening model, as well as numerical calculations of MCDF-SCFTIS \cite{Lixiangdong3} and AA-IS models \cite{chenzhanbin1,chenzhanbin2}.
(a) $1s2p-1s^2$ transition in Al$^{11+}$, (b) $1s3p-1s^2$ transition in Cl$^{15+}$. }\label{fig2}
\end{figure}

Based on equations (\ref{eq12}) and (\ref{eq14}), the line shifts of Al$^{11+} (1s2p-1s^2)$ and Cl$^{15+} (1s3p-1s^2)$ are computed and compared with experimental results \cite{Stillman,Beiersdorfer}, as well as the numerical calculations of multi-configuration-Dirac-Fock self-consistent finite-temperature ion-sphere model (MCDF-SCFTIS) \cite{Lixiangdong3} and  average-atom ion-sphere model (AA-IS) \cite{chenzhanbin1,chenzhanbin2}, as shown in figure \ref{fig2}.
It is found that there are full agreements between the present results and the two experiments of Al$^{11+}$ and Cl$^{15+}$. 
But for MCDF-SCFTIS and AA-IS models, the shifts of Al$^{11+} (1s2p-1s^2)$ transition are significantly underestimated. 
In the latest work of Li and Rosmej \cite{Lixiangdong3}, they proposed an analytical b-potential approach, based on MCDF-SCFTIS model and one more adjustable parameter $b$ to characterize the plasma-electron density. In order to match the measured results, $b=4$ and $b=2$ are adopted for Al$^{11+} (1s2p-1s^2)$ and Cl$^{15+} (1s3p-1s^2)$, respectively. 
Whereas, there is no adjustable parameter in the present model, and the plasma screening effects experienced are dependent on the specific bound state considered, as discussed in the below.

In the present model, the limitation  $\varepsilon_p>\varepsilon_b$ is used to guarantee the quantum degeneracy between the initial and recombined bound electrons, in which $\varepsilon_p$ is the energy of recombined bound electron and $\varepsilon_b$  is the outermost initial bound orbital energy as shown in equation (\ref{eq10}). Therefore, the limitations of $\varepsilon_p>\varepsilon_{3p}$ and $\varepsilon_p>\varepsilon_{2p}$ are applied to compute the line shift of Cl$^{15+} (1s3p-1s^2)$ and Al$^{11+} (1s2p-1s^2)$, respectively. In order to gain deeper insight into the quantum degeneracy effect, the $\delta \rho(r)$ with different limitations of the bound orbital energies ($\varepsilon_{2p}$ and $\varepsilon_{3p}$) for Al$^{11+}$ ion are presented and compared in figure \ref{fig1}(b). It is found that the $\delta \rho(r)$ with limiting energy $\varepsilon_{2p}$ are significantly larger than the one with limiting energy $\varepsilon_{3p}$. This indicates that the plasma-electron density distribution and its screening effect are sensitive to the specific bound state, which can also be verified by the good agreements with the experimental results  \cite{Stillman,Beiersdorfer}. 

\textbf{\textit{CMD simulations}} $-$ 
As discussed above, the electron density distributions of the targeted ions are sensitive with models applied. 
For further validation of the present model, the classical molecular dynamic (CMD) simulations of ultra-cold neutral plasmas (UNPs) are performed to obtain the electron density distribution. 
Regarding CMD method, it is well-known that the multi-particle interactions between ions and electrons can be treated well. 
And it has been extensively applied to treat the dynamical evolution and coupling properties of UNPs \cite{Killian,Pohl,Guo}.
For a typical UNP with $T_e$ of 5K, $\rho_e$ of $10^9\,/cm^3$ and $\Gamma_e \approx 0.5$, the electron De Broglie wavelength $(\lambda \approx 60 \, nm)$ is much less than the plasma Wigner-Seitz radius ($a_{WS}\approx6200\, nm$) and the quantum degeneracy effects are negligible. 
Therefore, UNPs can serve as a good prototype to study the many-body effects of moderately coupled plasmas using CMD method \cite{bergeson}. 
 
In the CMD simulations, 1000 electrons and 1000 singly charged ions are considered and the periodic boundary condition is applied to maintain constant density. 
With static screening approximation, the mobility of ions is neglected and the ions are fixed during the plasma evolutions. 
Then the electron density distribution induced by a given ion can be calculated by
\begin{equation}\label{eq15}
\delta \rho (r)= \rho (r)- \rho' (r),
\end{equation} 
where $\rho (r)$ and $\rho' (r)$ are the average electron densities obtained from two simulations of the UNP with and without this ion, respectively. The time step of $0.5 \,ps$ and the total evolution time of $25000 \, ps$ are employed for each simulation. For reducing the statistical error, these simulations are repeated 1000 times with different random initial positions and velocities of electrons. 

\begin{figure}[htbp]
\centering
\includegraphics[height=4.1cm,width=8.2cm]{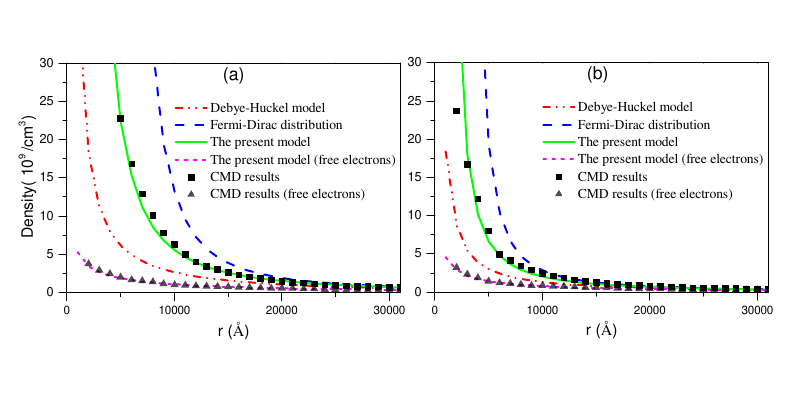}
\caption{ The electron density fluctuation $\delta \rho (r)$ around an ion ($Q$=1, at origin) embedded in the UNPs with (a) $\rho_e=10^9\, /cm^3,T_e=5.5\,K$ and (b) $\rho_e=10^9\, /cm^3,T_e=11.5\,K$. 
The distance $r$ is range from 0 to $a_{WS}/2\approx 31000$ \AA. The CMD simulated densities of electrons and free electrons are shown in \textbf{square} and \textbf{triangle}, respectively.
Electron densities from Fermi-Dirac distribution ($\textbf{dash line}$), linear Debye-H\" uckel model ($\textbf{dash-dot-dot line}$) and the present model ($\textbf{solid line}$) are shown to check their reliabilities. 
And the free-electron density from our model is shown in $\textbf{short dash line}$. }\label{fig3}
\end{figure} 

In order to further reveal the effect of recombined bound electron distribution, the free-electron density distributions of models are compared with CMD simulations. 
It is found that there are well agreements between the CMD and the present results, as well as the one of Fermi-Dirac distribution, which further indicate that the difference between full Fermi-Dirac distribution and CMD simulation only can come from the different treatments of negative-energy electrons distribution. 
Here, $p  = \sqrt{2m_e e\Phi_{DH}(\boldsymbol{r})}$ is approximated with Debye-Hückel potential $\Phi_{DH}(\boldsymbol{r})$. 
Therefore, the electron distribution cannot be described using equilibrated distribution theory, when the many-body effects become important for moderately/strongly coupled plasma.

\textbf{In conclusion},a new statistical model is proposed and successfully applied to treat the line shifts of  Cl$^{15+} (1s3p-1s^2)$ and Al$^{11+} (1s2p-1s^2)$ \cite{Stillman,Beiersdorfer}. 
In this model, the contributions of three-body recombination to the electron distributions are considered to describe the plasma-electron screening effect, which cause the electron distribution is far from equilibrium near the targeted ions in moderately/strongly coupled plasmas. 
The present model is further validated by the CMD simulations of UNPs' electron density distribution. 
Due to the quantum degeneracy effect between the recombined and the initial bound electrons, the electron distribution of the targeted ion is sensitive to specific bound state, which further influences the IPDs and line shifts.
The present model provides the basis for investigating atomic structure and dynamics processes in non-equilibrium dense plasmas, and then the opacity and EOS. 
In the next step, the present model will be validated for treating different plasma conditions, and machine leaning method will be devoted to construct a more convenient analytic potential, which is expected to provide a promising tool to handle those challenging questions existing in the related fields of high-energy-density physics, such as astrophysics, initial confinement fusion, and the study of matter in extreme conditions, and is also helpful for the understanding of quantum many-body interactions.

\section*{Acknowledgements} \label{sec:acknowledgements}
This work was supported by the National Key Research and Development Program of China under Grants No. 2017YFA0402300 and No. 2017YFA0403200 and the National Natural Science Foundation of China (Grants No.11474032, No.11774344, No.11704040, No.11534011 and No.U1530261). 

\bibliography{main}

\end{document}